# Control of ferromagnetism by manipulating the carrier wavefunction in ferromagnetic semiconductor (In,Fe)As quantum wells

Le Duc Anh, Pham Nam Hai and Masaaki Tanaka[a]

*Department of Electrical Engineering and Information Systems,*

*The University of Tokyo, 113-8656 Tokyo, Japan*

[a] Author to whom correspondence should be addressed.
Electronic mail: masaaki@ee.t.u-tokyo.ac.jp

We demonstrated the control of ferromagnetism in a surface quantum well containing a 5-nm-thick n-type ferromagnetic semiconductor (In,Fe)As layer sandwiched between two InAs layers, by manipulating the carrier wavefunction. The Curie temperature ($T_C$) of the (In,Fe)As layer was effectively changed by up to 12 K ($\Delta T_C/T_C = 55\%$). Our calculation using the mean-field Zener theory reveals an unexpectedly large *s-d* exchange interaction in (In,Fe)As. Our results establish an effective way to control the ferromagnetism in quantum heterostructures of n-type FMSs, as well as require reconsideration on the current understanding of the *s-d* exchange interaction in narrow gap FMSs.

Carrier-induced ferromagnetic semiconductors (FMSs) have nowadays become an important category of materials because they possess novel functionalities that cannot be achieved in conventional ferromagnetic metals, including the ability of controlling ferromagnetism through manipulation of carrier characteristics as well as band engineering in nanostructures. Particularly, "wavefunction engineering" of ferromagnetism by manipulating the spatial overlap of carrier wavefunctions and local magnetic moments has been proposed as an effective way to control the ferromagnetism in quantum well (QW) structures of FMSs[1,2]. On the other hand, the interplay of carriers and local magnetic moments in the "wavefunction engineering" manner gives insightful information about the physics of ferromagnetism of the FMS itself. However, such a novel "wavefunction engineering" of ferromagnetism has not been definitely demonstrated so far. In p-type FMSs such as (Ga,Mn)As, coherence of holes persists only at extremely low temperature (25 mK ~ 650 mK)[3]. Therefore, although there are reports on the modulation of ferromagnetism by electrical gating in field-effect transistor structures with (In,Mn)As[4] or (Ga,Mn)As[5,6,7] channels, these effects were induced by changing the local hole concentration, rather than by manipulating the quantized hole wavefunctions in the magnetic layers.

In this Letter, we demonstrate the intrinsic "wavefunction engineering" of ferromagnetism in surface tri-layer QWs consisting of an ultrathin (In,Fe)As layer sandwiched by two InAs thin layers. (In,Fe)As is a new n-type electron-induced FMS that has been successfully grown recently[8,9]. An important feature of this material is that the Fe atoms are in the neutral state ($Fe^{3+}$) when partially replacing the In atoms in the zinc-blende crystal structure. Therefore, the Fe atoms play only the role of local magnetic moments, neither of acceptors nor of donors. Electron carriers in (In,Fe)As can be supplied independently by co-doping with non-magnetic dopants, such as Be double-donors or Si donors. These electrons freely move in the conduction band rather than hop in the Fe-related impurity band, which is evident from their relatively high mobility (up to 160 cm$^2$/Vs), light effective mass (0.03 ~ 0.17 $m_0$), as well as very little $s$-$d$ scattering at the Fermi level[9,10]. (In,Fe)As exhibits an electron-induced ferromagnetism with a threshold electron density as low as 6×10$^{18}$ cm$^{-3}$, one order of magnitude smaller than that of holes in (Ga,Mn)As. In this study, we show that electron carriers in the tri-layer QWs have a long coherence length (above 40 nm) by observing the quantum-size effect (QSE), and that the wavefunction of electron carriers is not confined in the (In,Fe)As layer but extending to the surrounding InAs layers. Utilizing QSE, we manipulated the electron wavefunction in a 5 nm-thick InAs / 5 nm-thick (In,Fe)As / 5 nm-thick InAs QW by precise wet etching and successfully changed the Curie temperature ($T_C$) by 55% by removing only the nonmagnetic 5-nm-thick InAs surface layer.

Figure 1(a) shows the schematic sample structure studied here. The QW consists of InAs / Be doped (In$_{0.95}$,Fe$_{0.05}$)As / InAs tri-layers grown by low-temperature molecular beam epitaxy on semi-insulating GaAs (001) substrates[11]. Table I shows the structure parameters of sample A-F. All the samples except for sample A are ferromagnetic, and their $T_C$ values are shown in Table I. The conduction band of an InAs / GaAs heterojunction is known to have a large band offset (0.9 eV, 0.5 eV, 0.34 eV at the Γ, L, X point[12,13], respectively). This band offset, together with a large vacuum barrier at the surface of the top InAs layer[14], confines electrons inside the InAs / (In,Fe)As / InAs tri-layer QWs, and induces quantized electronic states in the conduction band of the QWs. The bulk-like sample F, which contains a 100 nm-thick (In,Fe)As layer grown on a GaAs substrate, is used as a reference.

To investigate QSE in our samples, we employed magnetic circular dichroism (MCD) spectroscopy in a reflection setup to probe the spin-polarized band structure of (In,Fe)As[11]. The MCD spectrum of the 100-nm-thick sample F (black curves in Fig 1(b) and 1(c)) indicates the bulk values of the optical transition energies $E_1$ (2.61 eV), $E_1+\Delta_1$ (2.88 eV), $E_0$' (4.39 eV), and $E_2$ (4.74 eV) of InAs. Figure 1(b) shows the MCD spectra of sample A, B, and C, which have fixed 5 nm-thick InAs top and bottom layers but different (In,Fe)As layer thickness $t$ of 2, 5 and 10 nm, respectively. MCD spectra of samples A, B and C exhibit the same features as those of sample F, but their optical transition energies show increasing blue shift from the bulk values with decreasing $t$. Note that the blue shift is as large as ~ 100 meV. This result clearly demonstrates the QSE in our QWs with (In,Fe)As layers. Next, we measured the MCD spectra of samples B, D, and E (Fig. 1(c)), which have the same top InAs layer thickness and $t$, but different bottom InAs layer thickness of 5, 20 and 30 nm corresponding to $L$ = 15, 30 and 40 nm, respectively, where $L$ is the total thickness of the tri-layer. Despite the fixed $t$, the MCD spectra of sample B, D and E again show systematic blue shift of the optical transition energies from the bulk values with decreasing $L$. This clearly indicates that QSE is determined not only by $t$ but also by $L$. The appearance of QSE in a 40 nm QW (sample E) indicates a long coherence length above 40 nm for electrons in our QWs. The extending wavefunction results from the fact that electron carriers are in the conduction band[9,10].

In order to systematically control QSE and perform wavefunction engineering of ferromagnetism, we gradually decreased the thickness of the tri-layer QW in sample B by precise wet etching[11]. Etching was conducted at a rate of 1.2 nm/step and repeated for 8 steps until the (In,Fe)As layer has been totally erased. From atomic force microscopy characterizations and MCD spectra analyses, we confirmed that the etching process was well-controlled, and the surface morphology and average roughness ($R$a) of sample B before and after etching are almost the same[11]. Figure 2(a) shows the MCD spectra of sample B after each step of etching, in which the increasing blue shift of $E_1$ and $E_0$' peaks of (In,Fe)As were observed clearly from the 1st step. As shown in Fig. 2(b), the intensity of peak $E_1$ remained almost constant until the 4th step of etching, and then linearly decreased down to zero after the 8th step. This behavior clearly indicates that the first four steps correspond to the etching of the top 5 nm-thick InAs layer, and the last four steps corresponds to the etching of the 5 nm-thick (In,Fe)As layer. As shown in Fig. 2(a), we fitted the whole MCD spectra near $E_1$ by Lorentzian curves, which gives the the peak position of the $E_1$ optical transition energy[11]. During the etching of the top InAs layer, the shift of $E_1$ transition energy from the bulk value increased from 70 meV to 126 meV (Fig. 2(c)). These results clearly indicate that the quantized electronic structure in the (In,Fe)As QW is well manipulated by etching, starting from the etching of the top InAs layer.

Next, we investigate the change of $T_C$ during the etching of sample B. Figure 2(d) shows $T_C$ of the middle (In,Fe)As layer at each etching time, estimated by the Arrot plot of the MCD intensity - magnetic field characteristics. It was found that $T_C$ of the (In,Fe)As layer started to decrease from the initial value of 22 K to 10 K ($\Delta T_C = 12$ K) after the etching of the 5 nm-thick InAs top layer. Because the magnetic (In,Fe)As layer remained untouched, this change of $T_C$ is caused by the reduction of the overlap between the electron wavefunctions and the localized Fe spins in the (In,Fe)As layer, as the electron wavefunctions are moved toward the outside of the (In,Fe)As layer. Note that the same $\Delta T_C$ in Mn-based FMSs requires a large change of hole concentration of ~ $10^{20}$ cm$^{-3}$, which was obtained only with intense electrical gating of 3 - 5 MV/cm[4-7]. Our results show that "wavefunction engineering" is a very effective method for controlling the ferromagnetism.

We performed a self-consistent calculation of the conduction-band potential profile and electron wavefunctions in sample B[11], and then calculated the Curie temperature straightforwardly by adopting the mean field model for FMS quantum wells[1,2]. The potential profile was calculated from the space charge potential, conduction band offset, and electron exchange-correlation potential. Spin polarization of (In,Fe)As was neglected in the calculation of the quantum well potential, quantized levels, and wavefunctions, because the MCD peak's blue shift can be obtained from the non spin-polarized band structure. Since the effective ionized impurity concentration $N_d$ in (In,Fe)As and the effective mass $m_\Lambda$ of electrons at the $\Lambda$ point are unknown, we treated them as fitting parameters that can be determined from the experimental blue shift of $E_1$.

Figure 3(c) shows the fitting results of the blue shift of $E_1$ which give $N_d = 1.5 \times 10^{19}$ cm$^{-3}$ and $m_\Lambda = 0.033 m_0$, where $m_0$ is the free electron mass. This $N_d$ value agrees well with the data in 100 nm-thick (In,Fe)As sample with the same growth condition and Be doping level[8]. The QW potential profile, two occupied quantized levels, and the distribution of the electron density along the growth direction are shown in Fig. 3(a). The evolution of QW potentials and electron distributions with etching are shown in Fig. 3(b). Our calculation confirmed the important effect of the thickness reduction of the top InAs layer: The electron wavefunctions are shifted toward the GaAs substrate due to the QW potential deformation, resulting in a monotonic reduction of the 2D electron density in the (In,Fe)As layer from $3.9 \times 10^{12}$ to $1.7 \times 10^{12}$ cm$^{-2}$. The change of $T_C$, $\Delta T_C = 12$K, with a change in 2D electron density of only $2.2 \times 10^{12}$ cm$^{-2}$ is among the largest in FMSs reported so far.

We estimated $T_C$ by using the mean-field model description developed for FMS QWs[1,2] which has been modified for n-type FMS:

$$T_C^{MF} = \frac{S(S+1)}{12} \frac{J_{sd}^2}{k_B} \frac{m^*}{\pi \hbar^2} N_{Fe} \sum_{E_i < E_F} \int_{(In,Fe)As(n>n_{th})} |\varphi_i(z)|^4 dz \qquad (1)$$

Here, $z$ is the growth direction, $S$ is the spin angular momentum of an Fe atom (=5/2), $m^*$ is the electron effective mass at the $\Gamma$ point, $k_B$ is the Boltzmann constant, $J_{sd}$ is the s-d exchange interaction constant, $N_{Fe}$ is the Fe atom density, and $\varphi_i(z)$ is the wavefunction of the $i$th occupied quantum level. The integral was carried out inside the (In,Fe)As layer where the local electron density is above $6.2 \times 10^{18}$ cm$^{-3}$, which is the threshold for electron-induced ferromagnetism found in our previous study[8,9]. Figure 3(d) shows our calculation results of $T_C^{MF}$, which quantitatively describe our experimental results. $J_{sd}$ was estimated to be 0.25 eVnm$^3$, corresponding to an exchange interaction energy $N_0\alpha$ of 4.5 eV. This large $N_0\alpha$ shows that the n-type FMS (In,Fe)As has high potential for a high-$T_C$ material having good compatibility with semiconductor device structures.

The $N_0\alpha$ estimated above is unexpectedly large. The *s-d* exchange interaction is usually thought to be only potential interaction, thus as small as 0.2 eV in II-VI magnetic semiconductors[15]. It has been believed to be difficult to enhance the *s-d* kinetic exchange interaction and to realize electron-induced ferromagnetism in III-V semiconductors. Although the use of ultrathin quantum wells was theoretically suggested to be a possible approach[16], the $|N_0\alpha|$ value experimentally obtained in (Ga,Mn)As quantum wells was found to be smaller than 0.2 eV[17]. Furthermore, the strength of the *s,p-d* exchange interaction in FMSs is generally thought to be small in narrow-gap host materials due to the longer bond length[18,19]. Our results thus require a reconsideration of the theoretical understanding of the *s-d* exchange interaction, especially in narrow-gap n-type electron-induced FMSs.

Here we point out that by considering the relative energy of the *d* orbital of a Fe atom and the conduction band bottom, this large $N_0\alpha$ may be possible. The *s-d* exchange interaction energy derived from the Kondo-like Anderson Hamiltonian[20] is given by

$$N_0\alpha = -2|V_{sd}|^2 \left( \frac{1}{E_C - \varepsilon_d} + \frac{1}{U - E_C + \varepsilon_d} \right)$$[21,22]. Here, $E_C$ is the energy at the conduction band bottom, $\varepsilon_d$ is the energy of the *d* states of Fe, $U$ is the Coulomb repulsion between opposite-spin electrons in a *d* state, and $V_{sd}$ is the *s-d* mixing potential. As discussed by Anderson[20], generally the *d* orbital of a magnetic impurity and the *s* orbital *of the neighboring atom* are not orthogonal, which generates a non-zero $V_{sd}$[23]. Therefore, if the Fe 3*d*-level overlaps with the conduction band bottom as illustrated in Fig. 4 which is likely to happen in the case of a narrow-gap host, the energy difference ($E_C - \varepsilon_d$) is small and

$N_0\alpha$ could be remarkably large. For a rough estimation, we take $|V_{sd}| = 1$ eV, $E_C - \varepsilon_d = 0.42$ eV (equal to the band gap of (In,Fe)As), and neglect the second term in the brackets, yielding $N_0\alpha = 4.8$ eV which is close to 4.5 eV estimated from Fig. 3(d). It is worth noting that for the bulk-like samples of (In,Fe)As, $N_0\alpha$ estimated by the mean field model is also unexpectedly large (2.8 eV)[9]. The reason that $N_0\alpha$ of the quantum well samples is larger than that of the bulk-like samples is possibly the enhancement of magnetic susceptibility due to the electron-electron interaction in low-dimensional FMS systems[24,25], which gives higher $T_C$ in FMS quantum wells. We also point out that such a strong *s-d* exchange interaction would challenge the validity of the mean-field model, and thus further non-perturbative theoretical approaches and experimental investigations on the electronic structure of (In,Fe)As should be carried out for better understanding of this material.

In summary, $T_C$ of a (In,Fe)As channel was effectively changed by up to 12 K ($\Delta T_C/T_C = 55\%$) by manipulating the carrier wavefunction in tri-layer QW structure. The evolution of $T_C$ is well described by the mean-field model. The large *s-d* exchange interaction in (In,Fe)As, whose origin remains to be elucidated, suggests a new approach towards high-$T_C$ FMS in which narrow-gap FMSs would play a more important role[26].

**Acknowledgments**

This work was partly supported by Grant-in-Aids for Scientific Research including the Specially Promoted Research, the Project for Developing Innovation Systems of MEXT, and the FIRST Program of JSPS.

**References**

[1]B. Lee, T. Jungwirth, A. H. MacDonald, *Semicond. Sci. Technol.* **17,** 393 (2002).

[2]B. Lee, T. Jungwirth, A. H. MacDonald, *Phys. Rev. B* **61,** 15606 (2000).

[3]L. Vila, R. Giraud, L. Thevenard, A. Lemaitre, F. Pierre, J. Dufouleur, D. Mailly, B. Barbara, G. Faini, *Phys. Rev. Lett.* **98**, 027204 (2007).

[4] H. Ohno, D. Chiba, F. Matsukura, T. Omiya, E. Abe, T. Dietl, Y. Ohno and K. Ohtani, *Nature* **408,** 944-946 (2000).

[5]D. Chiba, F. Matsukura, H. Ohno, *Appl. Phys. Lett* **89,** 162505 (2006).

[6] M. Sawicki, D. Chiba, A. Korbecka, Y. Nishitani, J. A. Majewski, F. Matsukura, T. Dietl & H. Ohno, *Nature Phys.* **6,** 22-25 (2010).

[7]A. M. Nazmul, S. Kobayashi, S. Sugahara, M. Tanaka, *Jpn. J. Appl. Phys.* **43,** L233-236 (2004).

[8]P. N. Hai, L. D. Anh, S. Mohan, T. Tamegai, M. Kodzuka, T. Ohkubo, K. Hono, M. Tanaka, *Appl. Phys. Lett* **101,** 182403 (2012).

[9]P. N. Hai, L. D. Anh, M. Tanaka, *Appl. Phys. Lett,* **101,** 252410 (2012).

[10]P. N. Hai, D. Sasaki, L. D. Anh, M. Tanaka, *Appl. Phys. Lett.* **100,** 262409 (2012).

[11]See supplementary material at [*URL will be inserted by AIP*] for further details on methods and supplementary data.

[12]Y. C. Ruan, W. Y. Ching, *J. Appl. Phys* **62,** 2885 (1987).

[13]W. Porod, D. K. Ferry, *Phys. Rev. B* **27,** 2587-2589 (1983).

[14]W. Melitz, J. Shen, S. Lee, J. S. Lee, A. C. Kummel, R. Droopad, E. T. Yu, , *J. Appl. Phys* **108,** 023711 (2010).

[15]J. K. Furdyna, *J. Appl. Phys* **64**, R29 (1988).

[16]G. M. Dalpian, S. H. Wei, *Phys. Rev. B* **73,** 245204 (2006).

[17]R. C. Myers, M. Poggio, N. P. Stern, A. C. Gossard, D. D. Awschalom, *Phys. Rev. Lett* **95,** 017204 (2005) . See also Erratum: *Phys. Rev. Lett.* **95**, 017204 (2005).

[18]T. Dietl, H. Ohno, F. Matsukura, *Phys. Rev. B* **63,** 195205 (2001).

[19]T. Dietl, H. Ohno, F. Matsukura, J. Cibert, D. Ferrand, *Science* **287,** 1019-1022 (2000).

[20]P. W. Anderson, *Phys. Rev* **124,** 1 (1964).

[21]J. R. Schrieffer and P. A. Wolff, *Phys. Rev.* 149, 491 (1966).

[22]A. K. Bhattacharjee, G. Fishman, B. Coqblin, *Physica* 117B&118B (1983).

[23]R. E. Walstedt, L. R. Walker, *Phys. Rev. B*, **11,** 9 (1975).

[24]T. Dietl, A. Haury, Y. Merle d'Aubigne, *Phys. Rev. B* **55**, R3347-R3350 (1997).

[25] A. Haury, A. Wasiela, A. Arnoult, J. Cibert, S. Tatarenko, T. Dietl, and Y. Merle d'Aubigné, *Phys. Rev. Lett* **79,** 511-514 (1997).

[26]A. Zunger, S. Lany, H. Raebiger, *Physics* **3,** 53 (2010).

**Figure captions:**

FIG 1. (a) Sample structure. (b) MCD spectra of sample A (purple curve), B (red curve), C (blue curve) with the same top and bottom InAs layer thickness (5 nm) but different (In,Fe)As thickness $t$. (c) MCD spectra of sample B, D (green curve), E (yellow curve) with the same top InAs and (In,Fe)As layer thickness (5 nm) but different total tri-layer thickness $L$. The spectrum of sample F (black curve) is shown as a reference. Spectra were vertically shifted and scaled as indicated for clarity. The black arrows point to the position of the $E_1$ and $E_0'$ peaks of the MCD spectra which were obtained by Lorentzian fitting, indicating the blue shift with decreasing $t$ and $L$. All the MCD spectra are measured at 5K and at a magnetic field of 1T applied normal to the film plane.

FIG 2. (a) MCD spectra measured at 5K, 1T of sample B with decreasing the total thickness $L$ of the QW by etching (8 times). MCD peaks $E_1$ and $E_0$' increasingly shifted to higher energy. $E_1$ and $E_1+\Delta_1$ peaks of GaAs were also observed but remained unchanged during etching. The black dashed curves are the Lorentzian fitting of the (In,Fe)As MCD spectra around the $E_1$ peak. The black arrows point to the position of the $E_1$ and $E_0'$ peaks which were obtained by Lorentzian fitting, indicating the blue shift. (b) Intensity of peak $E_1$ of (In,Fe)As was almost unchanged in the first 4 times (during the etching of 5 nm top InAs layer), but linearly decreased after then by further etching the (In,Fe)As layer. (c) Blue shift of the peak $E_1$ from the bulk value (sample F) with etching. (d) $T_C$ started to decrease even when only the top InAs layer was etched, due to the shift of the electron wavefunction towards the GaAs substrate.

FIG 3. (a) Calculation results of the conduction band potential profile at the Γ point which formed the QW in unetched sample B (thick blue curve). The energies and wavefunctions of the two occupied quantized levels and the local electron density distribution (thin blue curve)

are shown. (b) Evolution of the QW potential profile (thick curves) and the electron distribution (thin curves) with etching the top InAs layer. (c) Experimental blue shift of the $E_1$ peak (red points), and fitting (black curve) by the calculated first quantized level in the QW with the ionized donor density ($N_d$) and effective mass $m_\Lambda$ as parameters. (d) $T_C$ calculated by the mean-field-theory (blue curve) and the experimental results (red point).

FIG 4. Schematic band structure of (In,Fe)As. The overlap of the *d*-level of Fe impurities and the conduction band possibly induces the large *s-d* exchange interaction in (In,Fe)As.

Table I. Structure parameters and $T_C$ of InAs / (In,Fe)As / InAs tri-layer surface QWs. All the samples were grown on GaAs(001) semi-insulating substrates. The Fe concentration and the Be doping concentration in (In,Fe)As is fixed at 5% and $5\times10^{19}$ cm$^{-3}$, respectively.

| Sample | Top InAs (nm) | $(In_{0.95},Fe_{0.05})As$ $t$ (nm) | Bottom InAs (nm) | Total tri-layer $L$ (nm) | $T_C$ (K) |
|---|---|---|---|---|---|
| A | 5 | 2 | 5 | 12 | 0 |
| B | 5 | 5 | 5 | 15 | 22 |
| C | 5 | 10 | 5 | 20 | 17 |
| D | 5 | 5 | 20 | 30 | 30 |
| E | 5 | 5 | 30 | 40 | 30 |
| F (reference) | 10 | 100 | 20 | 130 | 34 |

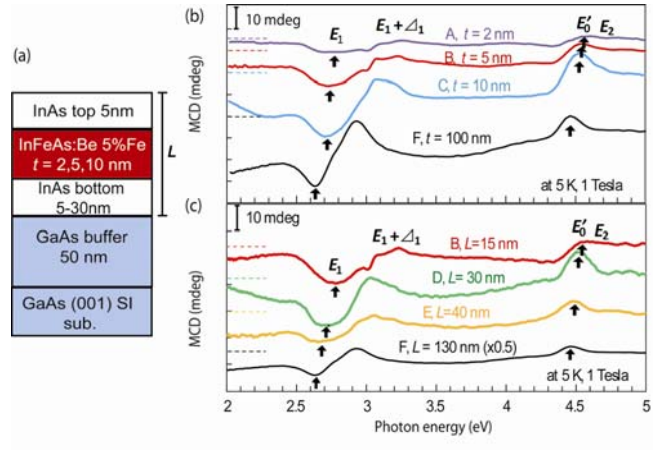

**FIG 1.** Anh *et al.*

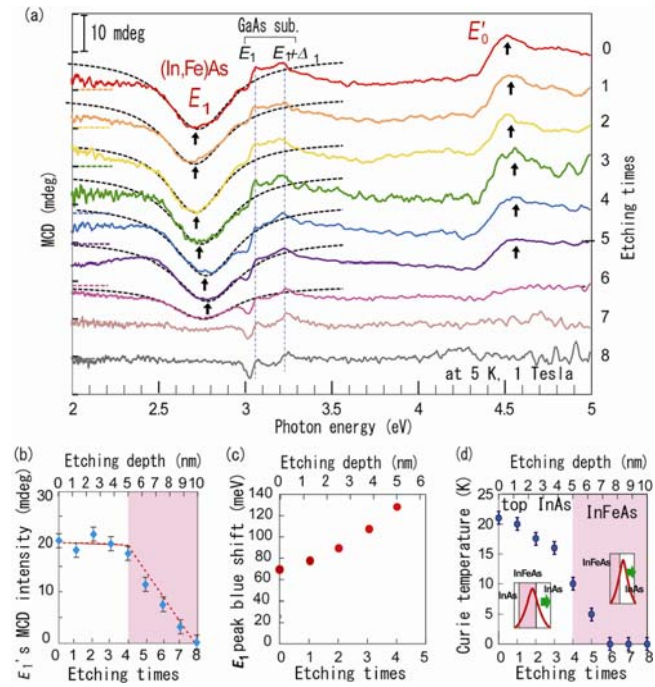

**FIG 2.** Anh *et al.*

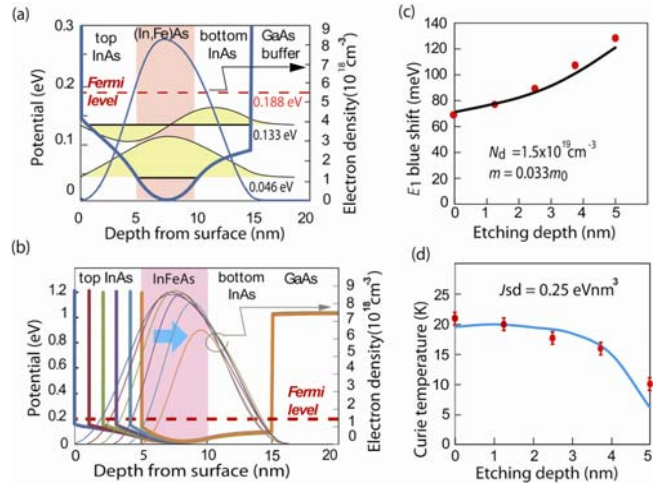

**FIG 3.** Anh *et al.*

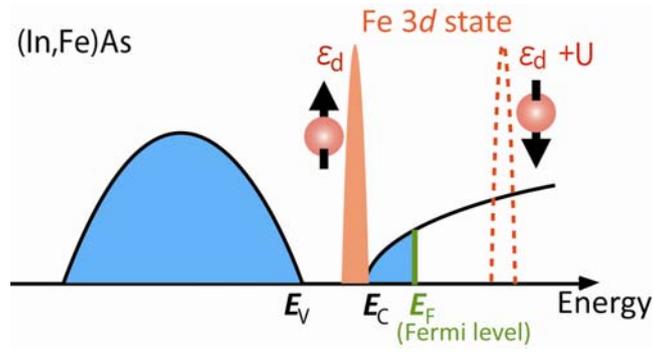

**FIG 4.** Anh *et al.*

# Supplemental Materials

## Control of ferromagnetism by manipulating the carrier wavefunction in ferromagnetic semiconductor (In,Fe)As quantum wells

Le Duc Anh, Pham Nam Hai, Masaaki Tanaka

*Department of Electrical Engineering and Information Systems,*

*The University of Tokyo, 7-3-1 Hongo, Bunkyo-ku, Tokyo 113-8656, Japan*

**Supplemental methods:**

**Sample preparation and characterization**

All the samples, whose structure is shown in Table 1 and Fig. 1(a) in the main text, were grown by molecular beam epitaxy (MBE) on semi-insulating GaAs (001) substrates. After growing a 50 nm-thick GaAs buffer layer at 580°C, we grew an InAs bottom layer (5-30 nm) at 500°C. The lattice mismatch between InAs and GaAs was quickly relaxed and a relatively smooth InAs surface was obtained, which was confirmed by *in-situ* reflection high energy electron diffraction (RHEED) patterns. After cooling the substrate temperature to 236°C, a Be doped $(In_{1-x},Fe_x)As$ thin film with $x = 0.05$ was grown with a thickness of 2, 5 and 10 nm. Finally, we grew a 5 nm-thick InAs top layer to complete the InAs / (In,Fe)As / InAs tri-layer QW structure. The Fe concentration was calibrated by secondary ion mass spectroscopy (SIMS). Be was co-doped in (In,Fe)As with a concentration of $5 \times 10^{19}$ cm$^{-3}$. Be atoms are considered to enter into the interstitial sites of (In,Fe)As due to the low growth temperature and become double-donors. The intrinsic and homogeneous ferromagnetism of (In,Fe)As samples grown with the same conditions has been confirmed elsewhere[1]. All the samples were cleaved into $1 \times 1$ cm$^2$ pieces for MCD measurements with reflection geometry.

**Reflection magnetic circular dichroism (MCD) and the Arrott plot**

In reflection MCD, we measure the difference in optical reflectivity between right ($R_{\sigma+}$) and left ($R_{\sigma-}$) circular polarizations, that is induced by the spin splitting of the band structure at a magnetic field $B$ of 1 T applied perpendicular to the film plane. The MCD intensity is expressed by $\mathrm{MCD} = \frac{90}{\pi} \frac{(R_{\sigma+} - R_{\sigma-})}{2} \sim \frac{90}{\pi} \frac{dR}{dE} \Delta E$, where $R$ is the optical reflectivity, $E$ is the photon energy, and $\Delta E$ is the spin-splitting energy (Zeeman energy) of the material. Because of the difference of the $dR/dE$ term, a MCD spectrum shows peaks corresponding to the optical transitions at critical point energies of the FMS band structure. At the same time, the MCD intensity is proportional to the magnetization $M$ of a material, because $M \propto \Delta E$ in FMSs. Therefore, MCD measurements give information of both the magnetization and the electronic structure of the material. Magnetization characteristics of all the samples were measured by magnetic field dependence of the MCD intensity at the $E_1$ peak of (In,Fe)As. $T_C$ of all the samples were estimated using the Arrott plot[2], which is based on the $MCD^2 - B/MCD$ plots at different temperatures. Figure S1 representatively shows the magnetization characteristics and the Arrott plots of sample B (5nm InAs/5 nm (In,Fe)As/5 nm InAs structure) (a) before and (b) after the etching of the 5 nm top InAs layer, respectively, measured by MCD. From the Arrott plots, the $T_C$ was estimated to be (a) 22 K and (b) 10 K, respectively.

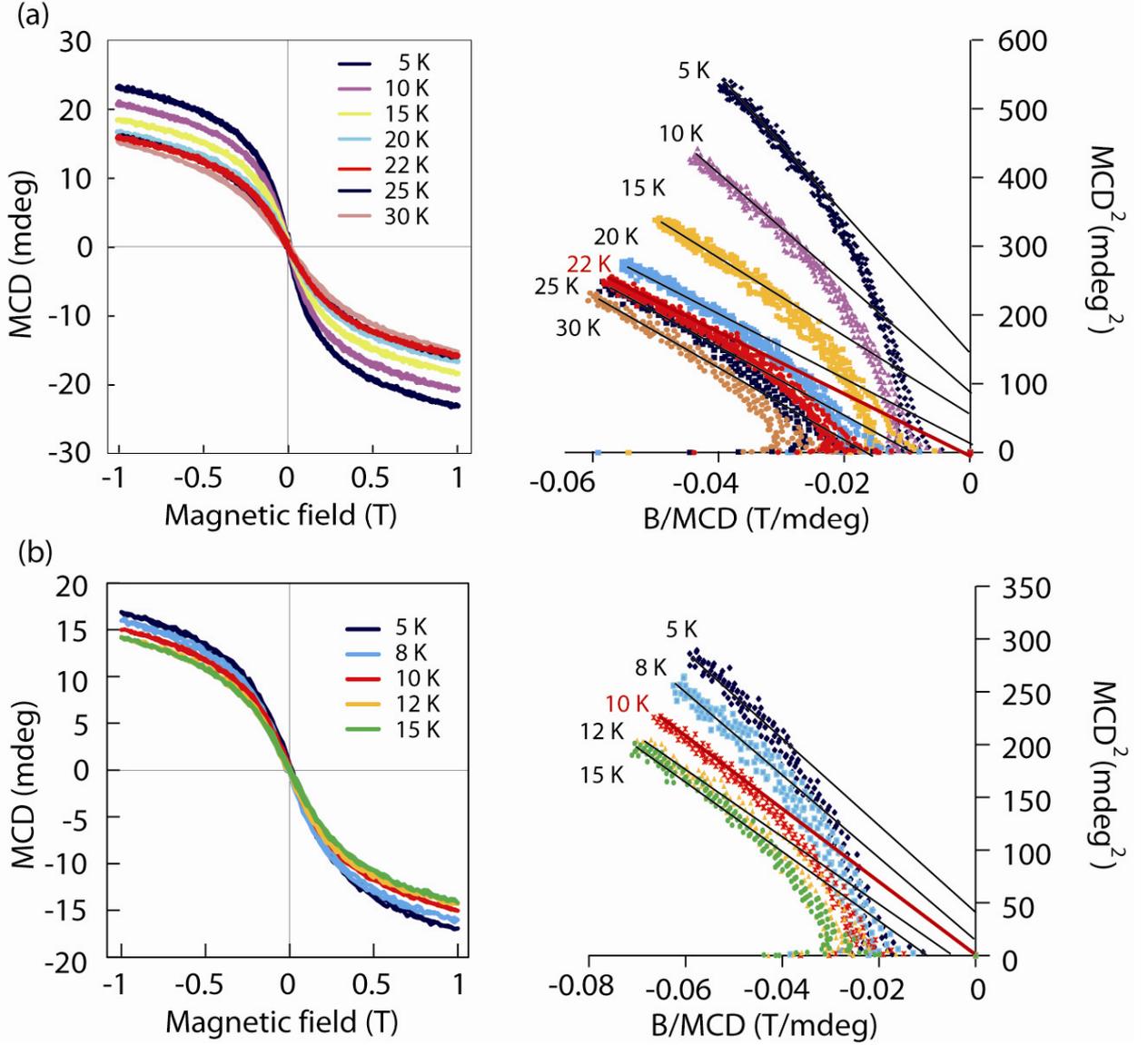

FIG. S1. Magnetization characteristics ($MCD$ vs. magnetic field $B$) and Arrot plots ($MCD^2$ vs. $B/MCD$) of sample B measured by MCD at $E_1$ peak: (a) before etching with $T_C$ of 22 K, (b) after etching 5 nm top InAs with $T_C$ of 10 K. The magnetic field was applied perpendicular to the film plane.

**Wet etching process**

We used diluted $H_3PO_4$+$H_2O_2$ solution as an etchant ($H_3PO_4$:$H_2O_2$:$H_2O$ = 0.25:0.25:100). During the etching, temperature was kept at 15$^o$C. We used this etchant for etching an InAs

(001) substrate for 1, 2, 3, 4, 5 minutes and then measured the etched depth by an XP-Plus 200 Surface Profilometer. We obtained a perfectly linear relation of the etching depth and the etching time, and estimated the etching rate to be 0.19 nm/s. Sample B was meticulously submerged in the etchant for 6 seconds each time (corresponding to an etched thickness of 1.2 nm), then its properties were characterized by MCD. The experiment was repeated until the MCD intensity of (In,Fe)As completely disappeared. Figure S2 shows the atomic force microscopy (AFM) images of a $100 \times 100$ nm$^2$ surface of sample B before and after the etching process. The average roughness values ($R$a) before and after etching were 0.404 and 0.345 nm, respectively, which are comparable to only one atomic layer of InAs (0.303 nm). This result indicates that the etching process was well-controlled and did not cause any inhomogeneous effect that could lead to the degradation of the magnetic layer. This conclusion is further supported by the analysis of MCD spectra mentioned below (Supplemental data 1).

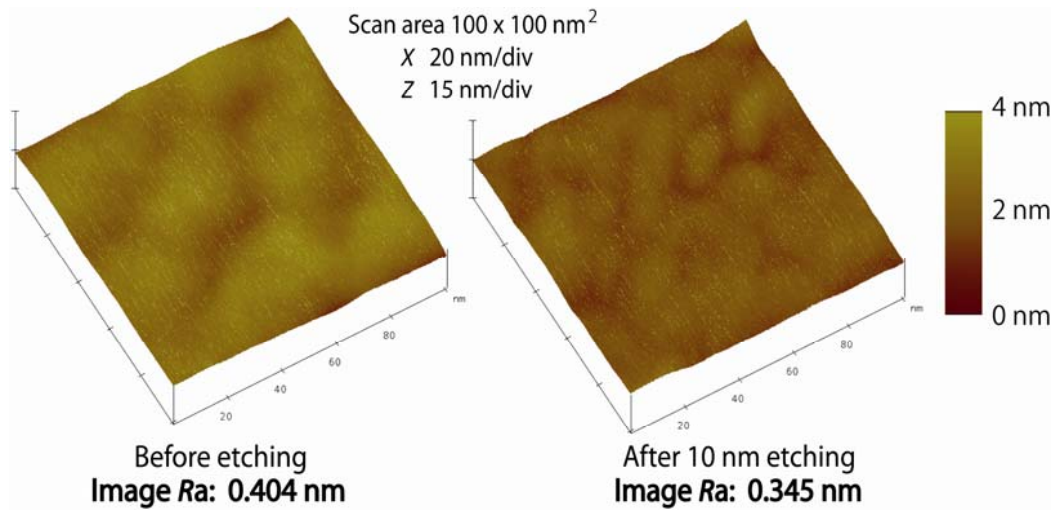

FIG. S2. Atomic force microscopy (AFM) images of sample B surface before and after 10 nm etching. The color bar indicates the $z$-range of 4 nm. The surface morphology and average roughness $R$a before and after etching is almost the same ($R$a values are 0.404 nm and 0.345 nm, respectively). Both $R$a values are comparable to only one atomic layer of InAs (0.303 nm). This result proves that the etching process was well-controlled and did not cause any apparent inhomogeneous effect on the quantum well that could lead to the degradation of the magnetic layer.

**Self-consistent calculation**

Self-consistent calculation of the electron wavefunctions and potential profile was performed neglecting the spin polarization of the conduction band. The calculation was based on the following equations:

$$\left(\frac{-\hbar^2}{2m^*}\frac{\partial^2}{\partial z^2} + V_{charge}(z) + V_{offset}(z) + V_{xc}(z)\right)\varphi(z) = E\varphi(z) \quad (S1)$$

$$\frac{\partial^2}{\partial z^2}V_{charge} = -e\frac{N_d(z) - \rho_e(z)}{\varepsilon} \quad (S2)$$

Here, $z$ is the growth direction, $V_{charge}$ is the space charge potential induced by ionized donors and electrons, $V_{offset}$ is the conduction band offset between InAs and GaAs at the Γ point, $V_{xc}$ is the exchange-correlation potential of electrons, and $\varphi(z)$ is the electron wavefunction. The electron density along the growth direction $\rho_e(z)$ was calculated as below, with the Fermi-Dirac distribution function at low temperature simplified as the Heaviside function:

$$\rho_e(z) = \sum_{E_i < E_F} \frac{m^*}{\pi\hbar^2}(E_F - E_i)|\varphi_i(z)|^2 \quad (S3)$$

The conduction band offset of InAs/GaAs at the Γ point is 0.9 eV (much larger than 0.17 eV for the valence band)[3]. The exchange-correlation potential was calculated from the local electron density based on the derivation of Gunnarsson and Lundqvist[4]. For simplicity, we took the barrier height at the InAs surface as infinity because of the large work function at the surface[5]. The space charge potential was calculated by Poisson's equation with the boundary condition that the Fermi level is pinned at 0.05 eV above the conduction band edge at the InAs surface[5,6] and at the mid-gap of the GaAs substrate. Ionized donors are thought to be uniformly distributed along the growth direction inside the (In,Fe)As layer, and its density $N_d$ was taken as a fitting parameter. The effective mass of electrons at the Γ point was set at 0.08 $m_0$, adopted from the data of the 100-nm-thick sample with the same order of carrier density ($1\times10^{19}$ cm$^{-3}$)[1]. For the dielectric constant of (In,Fe)As, the value of InAs $\varepsilon = 12.37\varepsilon_0$ was used. Note that the calculation neglecting the spin polarization still gives a good description of the blue shift in MCD spectra because MCD peaks appear at the critical point energies corresponding to the unpolarized band structure. Figure 3(a) of the main text shows the conduction band potential profile of the heterostructure which forms the QW, the electron wavefunctions of the 1st and 2nd quantized levels, as well as the electron density along the growth direction $\rho_e(z)$ calculated with fitted parameters of sample B.

The blue shift magnitude of $E_1$ peak is fitted by the calculated energy of the 1st quantized level in the tri-layer QW at the Λ point. Since the effective ionized impurity concentration $N_d$ in (In,Fe)As and the effective mass $m_\Lambda$ of electrons at the Λ point are unknown, we treated them as fitting parameters that can be determined from the experimental blue shift of $E_1$. Note that the fitting result is unique because the relative change of the blue shift during etching is dominantly determined by $N_d$, while the absolute magnitude of the blue shift is determined mainly by $m_\Lambda$. The quantized levels at the Λ point were calculated using the potential obtained from the above-mentioned calculation at the Γ point. Appropriate corrections were made for the band offset and the effective mass $m_\Lambda$ corresponding to the Λ point. Although the band offset at the Λ point is unknown, we assumed a band offset of 0.5 eV at the Λ point based on the existence of high barriers at the Γ (0.9 eV) and L (0.5 eV) points[3,7]. Note that a variation of the band offset at the Λ point from 0.5 ~ 0.9 eV does not affect much the calculation result of quantized level energies. The effective mass at the Λ point was obtained by fitting to the blue shift data of peak $E_1$. Here we neglected the hole energy quantization due to its heavy mass and low barrier height at the InAs/GaAs heterojunction in the valence band.

**Supplemental data**

1. **MCD spectra analysis of surface roughness**

 The conclusion from the AFM result in the main manuscript is further enforced by investigating the effect of etching on the width of the $E_1$ peak in the MCD spectra of sample B. In the following, we show that the dependence of the width of the $E_1$ peak on the quantum well thickness $L$ indicates that the surface roughness was indeed reduced with etching.

 As explained in the main manuscript, we fitted the whole MCD spectra near $E_1$ by Lorentzian curves, described by the following equation: $MCD = \dfrac{A}{\pi} \dfrac{\frac{\Delta}{2}}{(E-E_1)^2 + (\frac{\Delta}{2})^2}$, where $A$ is the MCD peak amplitude, $E_1$ is the peak position of the $E_1$ optical transition energy, $\Delta$ is the width of the $E_1$ peak. Here, $\Delta = \Delta_{bulk} + \delta\Delta(L)$, where $\Delta_{bulk}$ is the width of the $E_1$ peak for the bulk-like (In,Fe)As sample, and $\delta\Delta(L)$ is the broadening induced by the surface roughness at each quantum well (QW) thickness $L$ ($L$ was changed by etching). The fluctuation of $L$ due to the surface roughness causes fluctuation in quantized energies, leading to the broadening of the optical absorption peaks. Therefore, the surface roughness can be quantitatively evaluated by $\delta\Delta(L)$ as described below.

 Figures S3(a) and S3(b) show the MCD spectra and the fitting Lorentzian curves of bulk-like sample F and sample B (before etching), respectively. The $E_1$ peaks can be fitted very well with Lorentzian curves, giving experimental $E_1$ and $\Delta$ values. One can see that the width of the $E_1$ peak of sample B before etching ($\Delta = 0.46$ eV) is larger than that of bulk-like sample F ($\Delta_{bulk} = 0.29$ eV) as expected. Fig. S3(c) shows the experimental blue shift energy of the $E_1$ peak in sample B ($E_1$ - $E_{1bulk}$, where $E_{1bulk}$ is the $E_1$ peak energy of 100-nm-thick sample F) as a function of $L$. The relationship between ($E_1$ - $E_{1bulk}$) and $L$ can be empirically expressed by:

$$E_1 - E_{1bulk} = \frac{c}{L^{1.73}} \tag{S4}$$

Here $c$ is a constant. Note that in the case of ideal rectangular quantum wells with infinite barrier height, the relationship is $E_1 - E_{1bulk} = \frac{c}{L^2}$. From (S4), the fluctuation $\delta E_1$ of the quantized energy at $E_1$ can be related to the surface roughness $\delta L$ as: $\delta E_1 = \frac{1.73c}{L^{2.73}} \delta L$.

We then obtain

$$\frac{\delta E_1}{E_1 - E_{1bulk}} = \frac{1.73}{L} \delta L \tag{S5}$$

Since the MCD peak broadening $\delta\Delta(L) = \Delta - \Delta_{bulk}$ induced by the roughness is proportional to the fluctuation $\delta E_1$ of the quantized energy at $E_1$, we obtain the following relationship directly from equation (S5):

$$\frac{\Delta - \Delta_{bulk}}{E_1 - E_{1bulk}} \propto \frac{\delta L}{L} \tag{S6}$$

It is can be seen that the relationship (S6) is satisfied even in the case of ideal rectangular quantum wells with infinite barrier height.

Next, we assume that the surface roughness depends on $L$ as $\delta L \propto L^{\alpha}$. The physical meaning of $\alpha$ is that, in case of perfect homogeneous etching (in which the surface roughness does not change with etching), $\alpha=0$. If the surface roughness increases (decreases) with etching, we have $\alpha<0$ ($\alpha>0$). From equation (S6), $\alpha$ can be experimentally evaluated from the relationship:

$$\frac{\Delta - \Delta_{bulk}}{E_1 - E_{1bulk}} \propto \frac{\delta L}{L} \propto L^{\alpha-1} \tag{S7}$$

As clearly seen in Fig. S3(d), $\Delta$ remains almost constant during etching (as shown by the black dots, the $\Delta$ value increased slightly from 0.46 eV to 0.52 eV when $L$ is decreased from 15 nm to 10 nm). The data (black dots) in Figs. S3(c) and S3(d) give the relationship:

$$\frac{\Delta - \Delta_{bulk}}{E_1 - E_{1bulk}} \propto L^{0.92} \quad (\text{or } \alpha = 1.92)$$

The positive value of $\alpha$ indicates that the *surface roughness indeed decreased during etching*, which is consistent with the AFM result. These two pieces of evidence, AFM and MCD analyses, confirmed that our etching process is well-controlled and did not cause degradation in the (In,Fe)As magnetic layer.

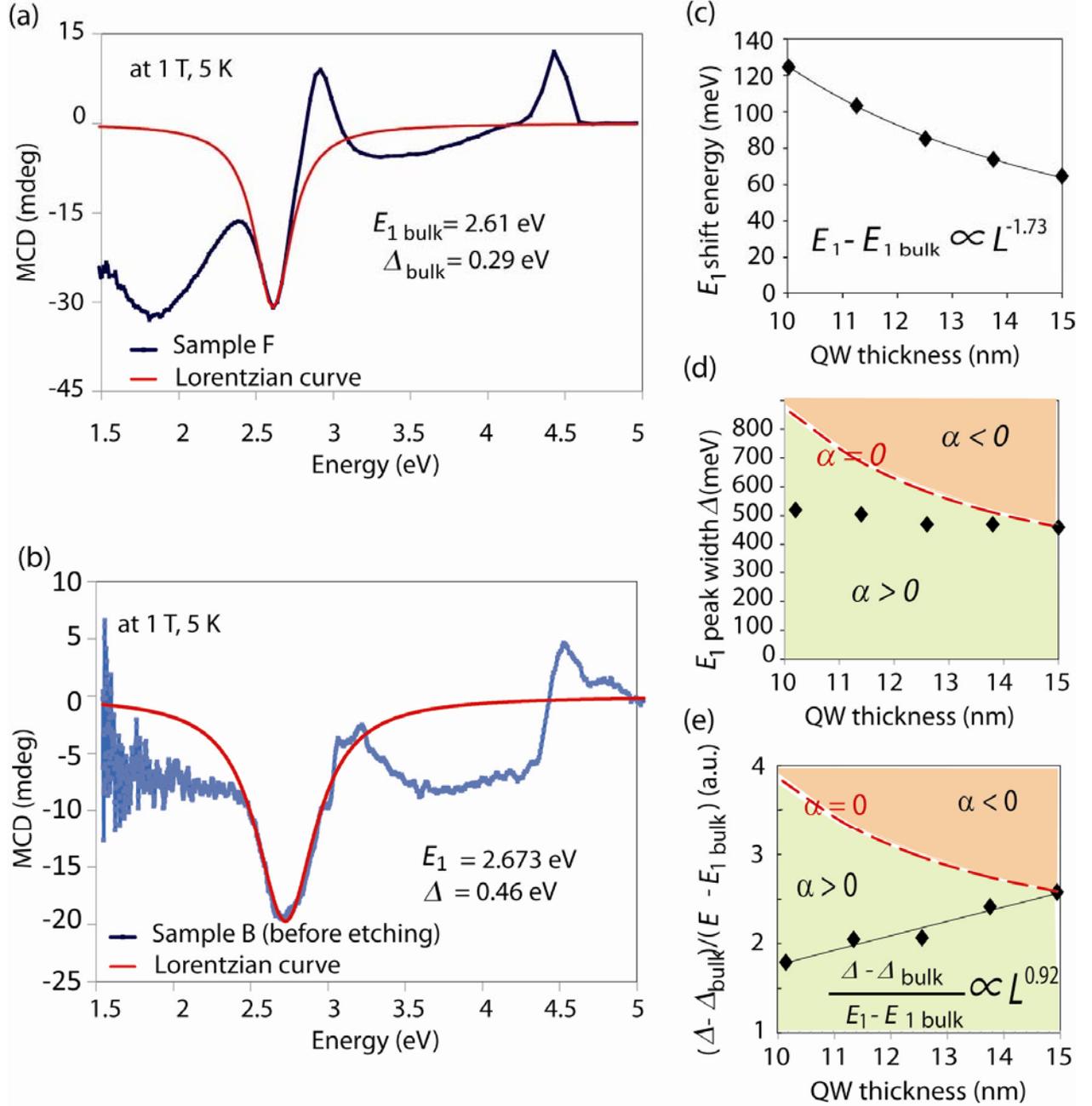

FIG. S3. Determination of the $E_1$ transition energy and peak width $\Delta$ with Lorentzian fitting in MCD spectra. (a) MCD spectra of 100-nm-thick (In,Fe)As (sample F) (blue curve) and its Lorentzian fitting (red curve). The fitted results are $E_{1\text{bulk}} = 2.61$ eV and $\Delta_{\text{bulk}} = 0.29$ eV. (b) MCD spectra of sample B before etching (blue curve) and its Lorentzian fitting (red curve).

The fitted results are $E_1 = 2.673$ eV and $\Delta = 0.46$ eV. (c) The experimental blue shift energy of the $E_1$ peak (i.e. $E_1 - E_{1\text{bulk}}$) in sample B. The dependence of the blue shift energy on the QW thickness $L$ can be deduced as $E_1 - E_{1\text{bulk}} \propto L^{-1.73}$ (the black solid curve). (d) Dependence of the width $\Delta$ of the $E_1$ peak on the QW thickness $L$ in sample B. $\Delta$ was almost constant when $L$ was decreased from 15 nm to 10 nm by etching. (e) The experimental results of the relationship in equation (1-4) that gives the positive $\alpha = 1.92$. This indicates that the surface roughness *decreased* with etching. Yellow areas, orange areas and the dashed red curves in (d) and (e) correspond to the case of positive $\alpha$ (average roughness decreased with etching), negative $\alpha$ (average roughness increased with etching), and $\alpha = 0$ (roughness did not change), respectively.

## 2. Full data of Fig. 3d in the main manuscript

We show in Fig. S4 the full data that are presented in Fig. 2d of our main manuscript. In Fig. S3, the blue circles are experimental $T_\text{C}$ values of sample B after each time of etching. The solid red curve is the calculated $T_\text{C}$ values based on the model described in the main manuscript. One can see that our mean field model calculation can well explain the decrease of $T_\text{C}$ up to the etching depth of 5 nm (i.e. the etching of the top 5 nm-thick InAs cap layer, corresponding to the white area in Fig. S4). When the middle (In,Fe)As layer was etched, however, the theoretical calculation underestimated $T_\text{C}$ (see the pink area in Fig. S4). *There are many possible reasons for this discrepancy.* We note that this underestimation of $T_\text{C}$ occurs only when the (In,Fe)As surface was exposed. Therefore, this may be due to the change of the pinning energy of the Fermi level at the (In,Fe)As surface compared with that of the InAs top layer. In Fig. S4, the red squares and red dashed curve are the calculated $T_\text{C}$ values of the etched (In,Fe)As based on the assumption that the Fermi level is pinned at 0.15 eV above the conduction band minimum on the (In,Fe)As surface rather than at 0.05 eV above the conduction band edge as in the case of InAs[6]. Although the calculated result can reasonably explain all the experimental $T_\text{C}$ values, we concentrate on *the effect of etching the top InAs layer*, while the middle 5 nm-thick (In,Fe)As remains untouched. This is because we want to avoid artifacts arising from many spurious effects when the (In,Fe)As layer was directly etched . This is one of the points which are different from other experiments that were conducted on (Ga,Mn)As, in which *the magnetic layer was etched directly*[8]. Thus we discussed only the $T_C$ values of sample B up to the fourth time of etching in the main manuscript.

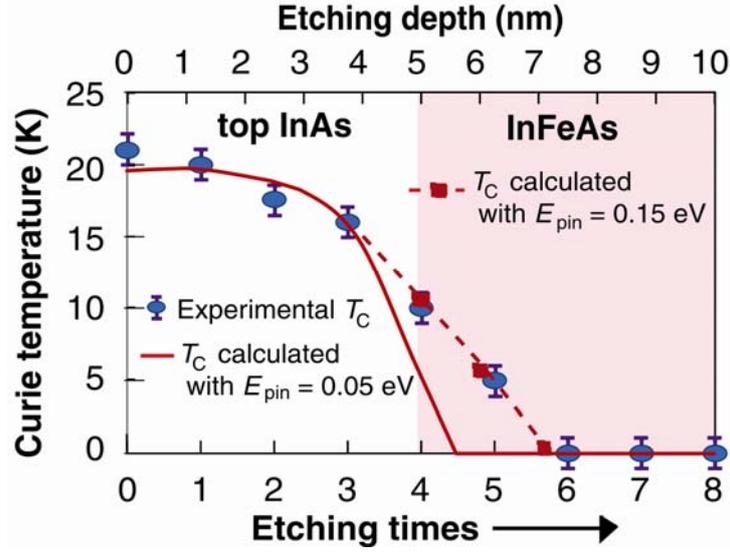

FIG. S4. Curie temperature ($T_C$) vs. etching times of sample B. This figure is the full data version of Fig. 3d, which contains all the data presented in Fig. 2d. The blue circles with error bars are experimental $T_C$ values of sample B after each time of etching. The solid red curve is the calculated $T_C$ values based on the mean field model described in the main manuscript. The red squares and the red dashed curve are the calculated $T_C$ values with the modified Fermi level's pinning energy ($E_{pin}$) at the (In,Fe)As surface. The calculation with $E_{pin}$ = 0.05 eV (solid red curve) seems to underestimate $T_C$ when etching into the (In,Fe)As:Be layer. These experimental $T_C$ values, however, can be fitted very well when we assumed $E_{pin}$ is changed to 0.15 eV (dashed red curve).

### 3. MCD spectra of a 20 nm-thick (In,Fe)As film grown on an InAs(001) substrate

We have grown a 20 nm-thick Be doped $(In_{0.95},Fe_{0.05})As$ on a undoped InAs (001) substrate with the same growth condition (sample G). The sample structure from the surface was Be doped (In,Fe)As 20 nm/InAs buffer 50 nm/InAs substrate. The Be doping concentration was $2.6 \times 10^{19}$ cm$^{-3}$. Excellent streaky RHEED patterns were observed during the MBE growth. $T_C$ of the sample was 13 K. Fig. S5 shows the MCD spectrum of this sample (red curve) measured at 5 K and 1 T. We also show the MCD spectra of two (In,Fe)As samples grown on semi-insulating GaAs (001) substrates for comparison: Bulk-like sample F ($t$ = 100 nm, $L$ = 130 nm, black curve) and sample C with a InAs/(In,Fe)As/InAs tri-layer QW ($t$ = 10 nm, $L$ = 20 nm, blue curve) grown on a GaAs buffer. The thickness of the (In,Fe)As layer of sample G is the same as $L$ = 20 nm of sample C. However, the MCD spectrum of sample G shows no blue shift relative to that of the bulk-like sample F, whereas we see a large blue shift in sample C. This clearly indicates that there is no significant potential barrier between the (In,Fe)As layer and the InAs buffer. As a result, electrons in (In,Fe)As extend deeply into the InAs buffer, which explains the relatively low $T_C$ of sample G.

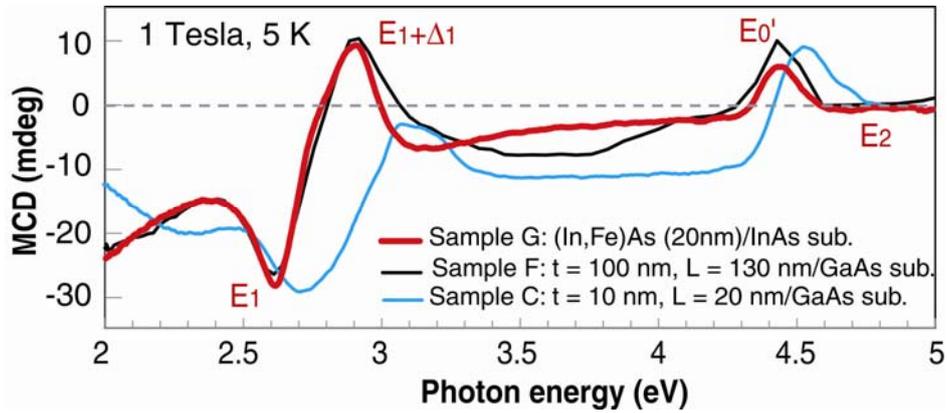

FIG. S5. MCD spectra of sample G with a 20-nm-thick (In,Fe)As layer grown on InAs substrate (red line). Bulk-like sample F with a InAs (10 nm)/(In,Fe)As (100 nm)/InAs (20 nm) tri-layer grown on a GaAs substrate (black line), and sample C with a InAs (5 nm)/(In,Fe)As (10 nm)/InAs (5 nm) tri-layer grown on a GaAs substrate (blue line) are also plotted for comparison. The MCD measurements were carried out in a reflection setup at 5 K at a magnetic field of 1 T applied perpendicular to the film plane. No blue shift was observed in sample G, indicating that there is no significant potential barrier between the (In,Fe)As layer and the InAs buffer.

**References**

[1] P. N. Hai, L. D. Anh, S. Mohan, T. Tamegai, M. Kodzuka, T. Ohkubo, K. Hono, M. Tanaka,, *Appl. Phys. Lett* **101,** 182403 (2012).

[2] Arrott, A., *Phys. Rev.* **108,** 1394 (1957).

[3] Y. C. Ruan, W. Y. Ching, *J. Appl. Phys* **62,** 2885 (1987).

[4] O. Gunnarsson, B. I. Lundqvist, *Phys. Rev. B* **13**, 4274-4298 (1976).

[5] W. Melitz *et al.*, *J. Appl. Phys* **108,** 023711 (2010).

[6] C. A. Mead, W. G. Spitzer, *Phys. Rev. Lett* **10,** 472 (1963).

[7] W. Porod, D. K. Ferry, *Phys. Rev. B* **27,** 2587-2589 (1983).

[8] O. Proselkov, D. Sztenkiel, W. Stefanowicz, M. Aleszkiewicz, J. Sadowski, T. Dietl, and M. Sawicki, *Appl. Phys. Lett* **100**, 262405 (2012).